\documentclass[twocolumn,english,notitlepage,superscriptaddress,nofootinbib]{revtex4-1}
\usepackage[T1]{fontenc}
\usepackage[latin9]{inputenc}
\usepackage{bm}
\usepackage{amsmath}
\usepackage{amssymb}

\makeatletter
\usepackage{babel}
\usepackage{footmisc}

\makeatother

\usepackage{babel}
\begin{document}
\title{Mechanism for the quantum natured gravitons to entangle masses}
\author{Sougato Bose}
\affiliation{Department of Physics and Astronomy, University College London, Gower
Street, WC1E 6BT London, UK}
\author{Anupam Mazumdar}
\affiliation{Van Swinderen Institute, University of Groningen, 9747 AG, The Netherlands}
\author{Martine Schut}
\affiliation{Van Swinderen Institute, University of Groningen, 9747 AG, The Netherlands}
\affiliation{Bernoulli Institute, University of Groningen, 9747 AG Groningen, the
Netherlands}
\author{Marko Toro\v{s}}
\affiliation{School of Physics and Astronomy, University of Glasgow, Glasgow, G12
8QQ, UK}
\begin{abstract}
This paper points out the importance of the quantum nature of the
gravitational interaction with matter in a linearized theory of quantum
gravity induced entanglement of masses (QGEM). We will show how the
quantum interaction entangles the steady states of a closed system
(eigenstates) of two test masses placed in the harmonic traps, and
how such a quantum matter-matter interaction emerges from an underlying
quantum gravitational field. We will rely upon quantum perturbation
theory highlighting the critical assumptions for generating a quantum
matter-matter interaction and showing that a classical gravitational
field does not render such an entanglement. We will consider two distinct
examples; one where the two harmonic oscillators are static and the
other where the harmonic oscillators are non-static. In both the cases
it is the quantum nature of the gravitons interacting with the harmonic
oscillators that are responsible for creating an entangled state with
the ground and the excited states of harmonic oscillators as the Schmidt
basis. We will compute the concurrence as a criterion for the above entanglement and highlight the role of the spin-2 nature of the graviton for entangling the two harmonic oscillators.

\end{abstract}
\maketitle

\section{Introduction}

The classical theory of general relativity (GR) is outstanding in
matching the observations on large scales, especially from the solar
system tests to the observations from the detection of the gravitational
waves~\citep{Will:2014kxa}. Despite these successes, the classical
theory fails at very short distances and early times. The classical
GR predicts black hole and cosmological singularity where the notion
of space-time breaks down~\citep{Hawking:1973uf}.

Although it is believed that the quantum theory of gravity will alleviate
some of these challenges, however, we still do not know whether gravity
is indeed quantum or not. Moreover, there are also many candidates
for a quantum theory of gravity~\citep{Kiefer}. From an effective
field theory perspective and at low energies, it is believed that
the gravitational interaction is being mediated by a massless spin-2
graviton, which can be canonically quantized~\citep{Gupta-1952,Gupta,Feynman,DeWitt}.
Although the perturbative quantum theory of gravity also possesses
many challenges, such as the issues of renormalisability at very high
energies and the issue of finiteness, at low energies where the day
to day experiments are performed, it is still a very good effective
field theory description of nature~\citep{Donoghue:1994dn}.

Given the feeble interaction strength of gravity, it is extremely
hard to detect a graviton in a detector by the momentum transfer~\citep{Dyson}.
Indirect detection of the quantum properties of the graviton remains
elusive in the primordial nature of the gravitational waves (GWs)~\citep{Ashoorioon,Venin}.
Astrophysical and cosmological uncertainties shroud any validation
of the quantum nature of space-time by modifying the photon dispersion
relationship~\citep{Addazi:2021xuf}. Moreover, the strict constraint
on the graviton mass indirectly arising from the propagation of the
GWs detected by the LIGO observatory hints no departure from GR in
the infrared~\citep{LIGOScientific:2016lio}.

Given all these challenges, it is worth asking how to test the quantum
nature of a graviton in a laboratory at low energies. Recently, there
has been a proposal to test the quantum nature of gravity by witnessing
the spin entanglement between the two quantum superposed test masses,
known as quantum gravity induced entanglement of masses (QGEM)~\citep{Bose:2017nin,Marshman:2019sne}.
The idea is to create a spatial quantum superposition of two test
masses and bring them adjacent to each other in a controlled environment
such that their only dominant interaction that remains is the exchange
of a massless graviton. It is possible to realise such a daunting
experiment but there are many challenges needed to be overcome\footnote{The detailed analysis of the demanding nature of the QGEM experiment
(such as creating Schr\"{o}dinger cat states with massive test masses
along with achieving the required coherence life time required to
detect the entanglement) has been discussed already in~\citep{Bose:2017nin}.
A related idea was also proposed in \citep{Marletto}. These initial
works~\citep{Bose:2017nin,Marshman:2019sne,Marletto} garnered extensive
interest in the research community \citep{Belenchia,Danielson,Nguyen,Pedernales,Marshman,vandeKamp:2020rqh,Toros,Kim,Toros:2020dbf,Christodoulou-1,Christodoulou-2,margalit,Tilly:2021qef,Carney,CarneyC,Streltsov1,Howl,Miki,Matsumura,Qvarfort,Miao,Schut:2021svd,Dutta,Krishnanda,Cosco,Weiss,Rijavec,Bin}. }.

In this paper we will review the conceptual underpinnings of the QGEM
mechanism. The entanglement of the two masses emerges from ``Local
Operation and Quantum Communication (LOQC)'' where as no entanglement
would occur by "Local Operations and Classical Communication (LOCC)"~\citep{Bennett}.
The LOCC principle states that the two quantum states cannot be entangled
via a classical channel if they were not entangled to begin with,
or entanglement cannot be increased by local operations and classical
communication. The classical communication is the critical ingredient
which can be put to test when it comes to graviton mediated interaction
between the two masses. If the graviton is quantum, it would mediate
the gravitational attraction between the two masses and it would also
entangle them, hence confirming the QGEM proposal~\citep{Bose:2017nin,Marshman:2019sne}.

One of the aims of the current paper is to sharpen the argument of
LOCC for the purpose of QGEM, and highlight the role of the quantum
nature of the interactions for entangling the two quantum systems.
We will use basic quantum mechanics and perturbation theory to show
how the perturbed wave functions of the matter systems become entangled
\textit{solely} by the virtue of the quantum natured graviton. We
will furthermore highlight the relevant degrees of freedom of the
graviton which interacts with the quantized matter, and are responsible
for the entanglement in both the static and in a non-static case.

We will study this problem in the number state basis of two harmonic
oscillator states, and we will show that the perturbed state is an
entangled state even at the \textit{first order} in a quantum perturbation
theory~\cite{Balasubramanian:2011wt}. The \emph{quantum} interaction
between the two matter systems emerges from the change in the graviton
vacuum energy due to the presence of the two quantum harmonic oscillators.
In the QFT community this is a well known way to understand how contact
interactions emerge, see~\cite{Zee1}. We will show that in a static
limit this change in the vacuum energy is the same as that of the
Newton's potential at the lowest order in the Newton's constant, which
appears at the \emph{second order} in the perturbation theory. Furthermore,
the Newtonian potential is the energy shift of the gravitational vacuum.
In this case the relevant gravitational degrees of freedom required
to be quantized is comprised of both the \textit{spin-2} and \textit{spin-0}
components~\cite{Gupta-1952,Marshman:2019sne}. A similar interpretation
applies to the non-static case, except there are some details in the
components of the graviton which will get modified.

In particular, if the matter is quantized then the energy shift in
the gravitational field becomes an \textit{operator valued} interaction.
Since we have the quantum superpositions for the matter systems --
then the energy shift in the gravitational field will not be a real
number, resulting in the gravitational field itself being a non-classical
entity.

We will calculate the \textit{concurrence}~\citep{Hill:1997pfa}
as a way to measure the entanglement between the two harmonic oscillators
and show that the concurrence is always positive for the quantum interaction
between the graviton and the matter states~\footnote{In this paper we will consider only pure states to highlight the conceptual
points, but the analysis could be readily extended to more realistic
situations with mixed states to account for the internal/external
noise sources and environmental decoherence.}~\footnote{The entanglement features of harmonic oscillators in presence of the
interaction are quite well-known in the quantum optics literature,
see for example~\cite{Milburn}. Typically, the quantum nature of
the photon plays the role of the quantum interaction. However, our
aim here is to concentrate on the quantum nature of the graviton,
especially highlighting the graviton's dynamical degrees of freedom
which are responsible for the quantum interaction in enabling the
entanglement feature of the quantum harmonic oscillators. These dynamical
degrees of freedom of the graviton are very different in nature compared
to the photon. }.

This paper is organised in the following way. We will first briefly
recap the known results, i.e. the two quantum harmonic oscillators
(Sec.~\ref{sec:Two-harmonic-traps}), and show how the quantum interaction
is responsible for generating the entanglement (Sec.~\ref{sec:Interaction-induces-entanglement}).
We will then quantify the degree of entanglement using concurrence
which we will compute using perturbation theory. We then discuss the
special case where the interaction potential is generated by the gravitational
field in the regime of weak gravity (Sec.~\ref{sec:Gravitational-field}).
In particular, we will first show how the $\hat{T}_{00}$ component
of the stress-energy tensor generates entanglement -- $\vert00\rangle$
and $\vert11\rangle$ are the Schmidt basis of the entangled state,
where $\vert nN\rangle\equiv\vert n\rangle\vert N\rangle$, and $\vert n\rangle$
($\vert N\rangle$) denote the number state of the first (second)
harmonic oscillator (Sec.~\ref{subsec:Entanglement-from-h00}). We
will then consider entanglement via graviton in the non-static case
(Sec.~\ref{subsec:Entanglement-from-hij}). We will find that the
$\hat{T}_{0i}$ components of the stress-energy generate a two-mode
squeezed state of the two harmonic oscillators (Sec.~\ref{case1}).
In addition, we will show that the $\hat{T}_{ij}$ components of the
stress-energy tensor (which give rise to the GWs) generate entanglement
-- $\vert00\rangle$ and $\vert22\rangle$ are the Schmidt basis
of the entangled state, in line with the quadrupole nature of the
gravitational radiation (Sec.~\ref{case2}). We will finally conclude
with the consequences for the classical/quantum communication (Sec.~\ref{sec:Discussion}).


\section{Two quantum harmonic oscillators\label{sec:Two-harmonic-traps} }

Let us consider the two matter systems, denoted by A and B, which
are placed in the harmonic traps located at $\pm d/2$. We suppose
that the harmonic oscillators are well-localised, such that 
\begin{equation}
\hat{x}_{A}=-\frac{d}{2}+\delta\hat{x}_{A},\qquad\hat{x}_{B}=\frac{d}{2}+\delta\hat{x}_{B},\label{eq:xy}
\end{equation}
where $\hat{x}_{A}$, $\hat{x}_{B}$ are the positions, and $\delta\hat{x}_{A}$,
$\delta\hat{x}_{B}$ denote small displacements from the equilibrium.
The usual Hamiltonian for the two harmonic oscillators is given by:
\begin{equation}
\hat{H}_{\text{matter}}=\frac{\hat{p}_{A}^{2}}{2m}+\frac{\hat{p}_{B}^{2}}{2m}+\frac{m\omega_{\text{m}}^{2}}{2}\delta\hat{x}_{A}^{2}+\frac{m\omega_{\text{m}}^{2}}{2}\delta\hat{x}_{B}^{2},\label{eq:matter}
\end{equation}
where $\hat{p}_{A}$, $\hat{p}_{B}$ are the conjugate momenta, and
$\omega_{\text{m}}$ is the harmonic frequency of the two traps (assumed
equal for the two particles for simplicity). We now introduce the
adimensional mode operators for the matter by writing 
\begin{alignat}{2}
\delta\hat{x}_{A} & =\sqrt{\frac{\hbar}{2m\omega_{\text{m}}}}(\hat{a}+\hat{a}^{\dagger}), & \delta\hat{x}_{B} & =\sqrt{\frac{\hbar}{2m\omega_{\text{m}}}}(\hat{b}+\hat{b}^{\dagger}),\label{eq:modesxy}\\
\hat{p}_{A} & =i\sqrt{\frac{\hbar m\omega_{\text{m}}}{2}}(\hat{a}^{\dagger}-\hat{a}), & \hat{p}_{B} & =i\sqrt{\frac{\hbar m\omega_{\text{m}}}{2}}(\hat{b}^{\dagger}-\hat{b}),\label{eq:modepxpy}
\end{alignat}
which satisfy the usual canonical commutation relationships (the only
nonzero commutators are given by $[a,~a^{\dagger}]=1$, and $[b,~b^{\dagger}]=1$).
Using this notation the Hamiltonian can be written succinctly as:
\begin{equation}
\hat{H}_{\text{matter}}=\hat{H}_{A}+\hat{H}_{B},\label{eq:matter2}
\end{equation}
where $\hat{H}_{A}=\hbar\omega_{\text{m}}\hat{a}^{\dagger}\hat{a}$
and $\hat{H}_{B}=\hbar\omega_{\text{m}}\hat{b}^{\dagger}\hat{b}$.
We will now want to investigate the steady-state when the system is
perturbed by an interaction Hamiltonian $H_{AB}$. In particular,
we will show that in general any \emph{quantum} interaction will entangle
the two harmonic oscillators.


\section{Quantum interaction induces entanglement\label{sec:Interaction-induces-entanglement}}

Let us assume that the initial state of the matter-system is given
by 
\begin{equation}
\vert\psi_{\text{i}}\rangle=\vert0\rangle_{A}\vert0\rangle_{B},
\end{equation}
where $\vert0\rangle_{A}$ ($\vert0\rangle_{B}$) denote the ground
state of the first (second) harmonic oscillator (in the following
we will omit the subscripts A, B for the states to ease the notation).
Suppose we now introduce an interaction potential $\lambda H_{AB}$
between the two matter systems, where $\lambda$ is a small bookkeeping
parameter. The perturbed state is given by: 
\begin{equation}
\vert\psi_{\text{f}}\rangle\equiv\frac{1}{\mathcal{\sqrt{N}}}\sum_{n,N}C_{nN}\vert n\rangle\vert N\rangle,\label{eq:final}
\end{equation}
where $\vert n\rangle$, $\vert N\rangle$ denote the number states,
and the overall normalisation is given by $\mathcal{N}=\sum_{n,N}\vert C_{nN}\vert^{2}$.
We have that $C_{00}\equiv1$ (coefficient of the unperturbed state),
while the other coefficients are given by 
\begin{alignat}{1}
C_{nN} & =\lambda\frac{\langle n\vert\langle N\vert\hat{H}_{AB}\vert0\rangle\vert0\rangle}{2E_{0}-E_{n}-E_{N}},\label{eq:coefficients}
\end{alignat}
where $E_{0}$ is the ground-state energy for the harmonic oscillators
(equal for the two harmonic oscillators as we have assumed the same
trap frequency), and $E_{n},~E_{N}$ denote the energies of the excited
states.

Here we note the role of $\hat{H}_{AB}$ being a quantum operator.
If $H_{AB}$ were \textit{classical}, it would have an associated
c-number (complex number), which would yield $\langle n\vert\langle N\vert H_{AB}\vert0\rangle\vert0\rangle=0$,
by virtue of the orthogonality of the ground and the excited states
($|0\rangle|0\rangle$ and $|n\rangle|N\rangle$) of the two quantum
harmonic oscillators, as $n,~N>0$ in Eq.~\eqref{eq:coefficients}
\footnote{Let us clarify what we mean by the Hamiltonian acting on a quantum
state in a Hilbert space, which is by definition an operator, to be
associated with a number. Essentially, we mean that it could (a) be
proportional to the identity operator multiplied by a number, or (b)
be something nontrivial, but acts on an eigenbasis. Our statement
above holds for both the definitions.}. By the same argument, interactions acting as operators on only one
of the two quantum systems (i.e., without products of operators acting
on the two matter systems) cannot entangle the two systems. It is
thus instructive to rewrite the state in Eq.~(\ref{eq:final}) in
the following way \cite{Balasubramanian:2011wt} 
\begin{alignat}{1}
\vert\psi_{\text{f}}\rangle\sim & (\vert0\rangle+\sum_{n>0}A_{n}\vert n\rangle)(\vert0\rangle+\sum_{N>0}B_{N}\vert N\rangle)\nonumber \\
 & +\sum_{n,N>0}(C_{nN}-A_{n}B_{N})\vert n\rangle\vert N\rangle,\label{eq:final2}
\end{alignat}
where $A_{n}\equiv C_{n0}$ and $B_{N}\equiv C_{0N}$. The first line
in Eq.~(\ref{eq:final2}) would yield a separable state, while the
second line is responsible for \emph{entanglement} of the two matter
systems (the $A_{n}$ and $B_{N}$ terms will not contribute to the entanglement at
first order in perturbation theory). We can already see the stark difference between the LOQC
and the LOCC~\footnote{ The above discussion, of course, relies on initially pure states
evolving unitarily under a fixed Hamiltonian so that they remain pure.
The general notion of LOCC~\cite{Bennett}, as used in quantum information
is broader, distinguishing entangled states from classically mutually
correlated states. The above discussion of Eq.~(\ref{eq:final2})
can, of course, be easily generalized to mixed states and probabilistic
operations (simply several repeats of our argument for different initial
states and different Hamiltonians with their corresponding probabilities). }. The non-trivial part of a LOQC mechanism is now encoded in the terms
of the interaction Hamiltonian $\hat{H}_{AB}$ producing the second
line in Eq.~(\ref{eq:final2}). On the other hand, a LOCC mechanism
could produce the first line of Eq.~(\ref{eq:final2}), but not the
second line, as a classical interaction cannot entangle the two the
quantum states if they were not entangled to begin with~\footnote{A similar discussion was first adopted in the momentum space entanglement
in a perturbative quantum field theory, Ref.~\cite{Balasubramanian:2011wt},
where they argued that the entanglement entropy of and mutual information
between subsets of field theoretic degrees of freedom at different
momentum scales are natural observables in quantum field theory. Here
we will compare the degree of entanglement by computing the \textit{concurrence},
see the discussion below.}.

To quantify the degree of entanglement we can compute the \textit{concurrence}~\citep{Hill:1997pfa,Rungta}:
\begin{equation}
C\equiv\sqrt{2(1-\text{tr}[\hat{\rho}_{A}^{2}])},\label{eq:concurrence}
\end{equation}
where $\hat{\rho}_{A}$ can be computed by tracing away the B state
\begin{equation}
\hat{\rho}_{A}=\sum_{N}\langle N\vert\psi_{\text{f}}\rangle\langle\psi_{\text{f}}\vert N\rangle.\label{eq:rhoA}
\end{equation}
We will recall that the larger the concurrence $C$ is, the larger
is the degree of entanglement -- $C=0$ corresponds to a separable
state, while $C=\sqrt{2}$ is obtained for a maximally entangled state.
Inserting Eq.~(\ref{eq:final}) into Eq.~(\ref{eq:rhoA}) we find
\begin{equation}
\hat{\rho}_{A}=\frac{1}{\mathcal{N}}\sum_{n,n',N}C_{nN}C_{n'N}^{*}\vert n\rangle\langle n'\vert.\label{eq:rhoa}
\end{equation}
We will then insert Eq.~(\ref{eq:rhoA}) back into Eq.~(\ref{eq:concurrence})
to eventually find: 
\begin{equation}
C\equiv\sqrt{2(1-\sum_{n,n',N,N'}C_{nN}C_{n'N}^{*}C_{n'N'}C_{nN'}^{*}/\mathcal{N}^{2})}.\label{eq:concurrence2}
\end{equation}
In the next sections, we will consider the entanglement of two harmonic
oscillators induced by the quantum nature of gravitons. For this case,
the entanglement will be induced by the terms $C_{11}$ and $C_{22}$
at the lowest order in the perturbation theory when the potential
$\hat{H}_{AB}$ is generated by the quantized gravitational field
in the regime of weak gravity.


\section{Quantum Gravitational interaction\label{sec:Gravitational-field}}

We will consider the setup of two quantum harmonic oscillators (introduced
in the previous sections) in the presence of the gravitational field.
In particular, we will work in the regime of small perturbations $|h_{\mu\nu}|\ll1$
about the Minkowski background $\eta_{\mu\nu}$. The metric is given
by: $g_{\mu\nu}=\eta_{\mu\nu}+h_{\mu\nu}$ (where $\mu,\nu=0,1,2,3$
and we are using $(-,+,+,+)$ signature throughout). We will promote
the fluctuations into the quantum operators, 
\begin{equation}
\hat{h}_{\mu\nu}=\mathcal{A}\int d\bm{k}\sqrt{\frac{\hbar}{2\omega_{\bm{k}}(2\pi)^{3}}}(\hat{P}_{\mu\nu}^{\dagger}(\bm{k)}e^{-i\bm{k}\cdot\bm{r}}+\text{H.c}),\label{eq:hmunu}
\end{equation}
where ${\bm{k}}$ is the three-vector, and $d{\bm{k}}\equiv d^{3}k$.
The prefactor is denoted by $\mathcal{A}=\sqrt{16\pi G/c^{2}}$, where
$G$ is the Newton's constant, and $\hat{P}_{\mu\nu}$ and $\hat{P}_{\mu\nu}^{\dagger}$
denote the graviton annihilation and the creation operator. We will
discuss in detail the properties of the graviton and the relevant
degrees of freedom below.

Around the Minkowski background, the graviton coupling to the stress-energy
tensor $\hat{T}_{\mu\nu}$ is given by the following \textit{operator
valued} interaction term: 
\begin{equation}
\hat{H}_{\text{int}}=-\frac{1}{2}\int d\bm{r}\hat{h}^{\mu\nu}(\bm{r})\hat{T}_{\mu\nu}(\bm{r}),\label{eq:matter-gravity}
\end{equation}
where ${\bm{r}}$ denotes the 3-vector.

We will now consider separately the coupling induced by the component
$\hat{T}_{00}$ in the static limit and by the full stress-energy
tensor $\hat{T}_{\mu\nu}$ in the non-static case.


\section{Entanglement via graviton in the static limit \label{subsec:Entanglement-from-h00}}

Let us consider two particles of mass $m$ (which will form the two
oscillating systems). The two particles are generating the following
current in the static limit: 
\begin{equation}
\hat{T}_{00}(\bm{r})\equiv mc^{2}(\delta(\bm{r}-\hat{\bm{r}}_{A})+\delta(\bm{r}-\hat{\bm{r}}_{B})),\label{eq:current}
\end{equation}
where $\hat{\bm{r}}_{A}=(\hat{x}_{A},0,0)$, $\hat{\bm{r}}_{B}=(\hat{x}_{B},0,0)$
denote the positions of the two matter systems. The Fourier transform
of the current is given by 
\begin{equation}
\hat{T}_{00}(\bm{k})=\frac{mc^{2}}{\sqrt{(2\pi)^{3}}}(e^{i\bm{k}\cdot\hat{\bm{r}}_{A}}+e^{i\bm{k}\cdot\hat{\bm{r}}_{B}}),\label{eq:sources}
\end{equation}
where ${\bm{k}}$ denotes 3-momentum.

Following the canonical quantisation of graviton in a weak field regime~\citep{Gupta-1952},
we decompose $\hat{h}_{\mu\nu}=\hat{\gamma}_{\mu\nu}-(1/2)\eta_{\mu\nu}\hat{\gamma}$
around a Minkowski background (where we use the convention $\gamma\equiv\eta_{\mu\nu}\gamma^{\mu\nu}$).
The two distinct modes, i.e. the spin-2, $\gamma_{\mu\nu}$, and the
spin-0, $\gamma$, can be treated as independent variables. They are
promoted as self-adjoint operators, and decomposed into: 
\begin{alignat}{1}
\hat{\gamma}_{\mu\nu} & =\mathcal{A}\int d\bm{k}\sqrt{\frac{\hbar}{2\omega_{\bm{k}}(2\pi)^{3}}}(\hat{P}_{\mu\nu}^{\dagger}(\bm{k)}e^{-i\bm{k}\cdot\bm{r}}+\text{H.c}),\label{spin-2}\\
\hat{\gamma} & =2\mathcal{A}\int d\bm{k}\sqrt{\frac{\hbar}{2\omega_{\bm{k}}(2\pi)^{3}}}(\hat{P}^{\dagger}(\bm{k)}e^{-i\bm{k}\cdot\bm{r}}+\text{H.c}),\label{spin-0}
\end{alignat}
where 
\begin{alignat}{1}
\left[\hat{P}_{\mu\nu}(\bm{k}),~\hat{P}_{\lambda\rho}^{\dagger}(\bm{k}')\right] & =[\eta_{\mu\lambda}\eta_{\nu\rho}+\eta_{\mu\rho}\eta_{\nu\lambda}]\delta(\bm{k}-\bm{k}'),\label{eq:comm1}\\
\left[\hat{P}(\bm{k}),~\hat{P}^{\dagger}(\bm{k}')\right] & =-\delta(\bm{k}-\bm{k}').\label{eq:comm2}
\end{alignat}
The graviton Hamiltonian is now given by~\cite{Gupta-1952}: 
\begin{equation}
\hat{H}_{g}=\int d\bm{k}\hbar\omega_{\bm{k}}\left(\frac{1}{2}\hat{P}_{\mu\nu}^{\dagger}(\bm{k})\hat{P}^{\mu\nu}(\bm{k})-\hat{P}^{\dagger}(\bm{k})\hat{P}(\bm{k})\right)\,.
\end{equation}
We are interested in computing the change in the energy $\Delta{\hat{H}_{g}}$-
\emph{the shift of the energy of the graviton vacuum arising from
the interaction with the matter}. In the static limit (where we neglect
the motion of the two harmonic oscillators), the interaction Hamiltonian
can be written in a simple form: 
\begin{equation}
\hat{H}_{\text{int}}=\frac{1}{2}\int d{\bm{r}}\left[\hat{\gamma}_{00}({\bm{r}})+(1/2)\hat{\gamma}({\bm{r}})\right]\hat{T}_{00}({\bm{r}}).\label{eq:H00}
\end{equation}
We can now compute the shift to the energy of the graviton vacuum
using the perturbation theory. The first order term vanishes~\footnote{The first order contribution to the energy is given by $\langle0|\hat{H}_{\text{int}}|0\rangle=0$,
where $|0\rangle$ denotes the unperturbed graviton vacuum. This is
due to the fact that $\hat{H}_{\text{int}}$ depends linearly on $\hat{\gamma}_{\mu\nu},~\hat{\gamma}$
which are themselves linear combinations of creation and the annihilation
operators, $\hat{P}_{\mu\nu}^{\dagger},\hat{P}_{\mu\nu},\hat{P}^{\dagger},~\hat{P}$.
Hence $\langle0|\hat{H}_{\text{int}}|0\rangle$ depends only linearly
on $\hat{P}_{\mu\nu}^{\dagger},\hat{P}_{\mu\nu},\hat{P}^{\dagger},~\hat{P}$
and thus vanishes (as $\hat{P}\vert0\rangle=0$ and $\langle0\vert\hat{P}^{\dagger}=0$
and similarly for the other operators). The non-vanishing contribution
will come from the second order term in the perturbation theory~\citep{Cohen,Scadron,Marshman:2019sne}.\label{fn3}}, while the second order term in the perturbation theory yields: 
\begin{equation}
\Delta{\hat{H}_{g}}\equiv\int d\bm{k}\frac{\langle0\vert\hat{H}_{\text{int}}\vert\bm{k}\rangle\langle\bm{k}\vert\hat{H}_{\text{int}}\vert0\rangle}{E_{0}-E_{\bm{k}}},\label{eq:dH}
\end{equation}
where $\vert\bm{k}\rangle=(\hat{P}_{00}^{\dagger}(\bm{k})+\hat{P}^{\dagger}(\bm{k}))\vert0\rangle$
is the one particle state constructed in the unperturbed vacuum, $E_{\bm{k}}=E_{0}+\hbar\omega_{\bm{k}}$
is the energy of the one-particle state, and $E_{0}$ is the energy
of the vacuum state. The mediated graviton is now off-shell/virtual
by virtue of the integration of all possible momentum $\bm{k}$ --
and hence does not obey classical equations of motions. Using Eqs.~(\ref{eq:hmunu},
\ref{spin-2}, \ref{spin-0}) and (\ref{eq:H00}) we readily find~\footnote{Inserting the definition of $\hat{H}_{\text{int}}$ from Eq.~\eqref{eq:H00}
(and the definitions of $\hat{\gamma}_{\mu\nu}$ and $\hat{\gamma}$
from Eqs.~\eqref{spin-2} and \eqref{spin-0}, respectively) we encounter
the following expression $\langle\bm{k}\vert(\hat{P}_{00}^{\dagger}(\bm{k}')+\hat{P}^{\dagger}(\bm{k'}))\vert0\rangle.$
Using $\vert\bm{k}\rangle=(\hat{P}_{00}^{\dagger}(\bm{k})+\hat{P}^{\dagger}(\bm{k}))\vert0\rangle$
we then find 
\[
\langle0\vert(\hat{P}_{00}(\bm{k})\hat{P}_{00}^{\dagger}(\bm{k}')+\hat{P}(\bm{k})\hat{P}^{\dagger}(\bm{k'}))\vert0\rangle,
\]
while the other terms vanish as the vacuum state satisfies $\hat{P}_{00}(\bm{k})\vert0\rangle=\hat{P}(\bm{k})\vert0\rangle=0$.
The two terms on the right-hand side can then be rewritten as $\langle0\vert[\hat{P}_{00}(\bm{k}),\hat{P}_{00}^{\dagger}(\bm{k}')]\vert0\rangle$
and $\langle0\vert[\hat{P}(\bm{k}),\hat{P}^{\dagger}(\bm{k'})]\vert0\rangle$,
where we have used the definition of the commutator $[\hat{O}_{1},\hat{O}_{2}]=\hat{O}_{1}\hat{O}_{2}-\hat{O}_{2}\hat{O}_{1}$
(as well as again the definition of the of the vacuum state). Using
now the commutation relations defined in Eq.~\eqref{eq:comm1} and
\eqref{eq:comm2}, and summing the two terms we then finally obtain
\[
\langle\bm{k}\vert(\hat{P}_{00}^{\dagger}(\bm{k}')+\hat{P}^{\dagger}(\bm{k'}))\vert0\rangle=\delta(\bm{k}-\bm{k}').
\]
} 
\begin{equation}
\langle\bm{k}\vert\hat{H}_{\text{int}}\vert0\rangle=\frac{\mathcal{A}}{2}\sqrt{\frac{\hbar}{2\omega_{\bm{k}}}}\hat{T}_{00}(\bm{k}),\label{eq:hmunu-1}
\end{equation}
where we have used the definition of the Fourier transform 
\begin{equation}
\hat{T}_{00}(\bm{k})=\sqrt{\frac{1}{(2\pi)^{3}}}\int d\bm{r}e^{-i\bm{k}\cdot\bm{r}}\hat{T}_{00}(\bm{r}).\label{eq:T00k}
\end{equation}
From Eq.~\eqref{eq:hmunu-1} we then obtain a simple expression 
\begin{equation}
\langle0\vert\hat{H}_{\text{int}}\vert\bm{k}\rangle\langle\bm{k}\vert\hat{H}_{\text{int}}\vert0\rangle=\frac{\hbar\mathcal{A}^{2}\hat{T}_{00}^{\dagger}(\bm{k})\hat{T}_{00}(\bm{k})}{8\omega_{\bm{k}}}.\label{eq:unknown}
\end{equation}
From Eq.~(\ref{eq:dH}) we then readily find: 
\begin{equation}
\Delta{\hat{H}_{g}}=-\mathcal{A}^{2}\int d\bm{k}\frac{\hat{T}_{00}^{\dagger}(\bm{k})\hat{T}_{00}(\bm{k})}{8c^{2}\bm{k}^{2}},\label{eq:dH2}
\end{equation}
Performing the momentum integration using spherical coordinates we
then find the result 
\begin{equation}
\Delta{\hat{H}_{g}}=-\frac{\mathcal{A}^{2}m^{2}c^{2}}{16\pi\vert{\hat{\bm{r}}}_{A}-{\hat{\bm{r}}}_{B}\vert},\label{eq:dH2final}
\end{equation}
where we have omitted the self-energy terms of the individual particles~\footnote{There are self-energy contributions which provide the ultraviolet
(UV) corrections and tend to generate infinities in the limit when
the graviton momentum goes to infinity, i.e. $k\rightarrow\infty$.
This is an example of a UV divergence appearing in a perturbative
quantum gravity. We are interested in the infrared (IR) limit where
we are neglecting the UV aspects of the quantum gravity.} . We will finally insert $\mathcal{A}=\sqrt{16\pi G/c^{2}}$ into
Eq.~(\ref{eq:dH2final}) to find Newton's potential\footnote{There is a covariant formulation also to obtain the same answer by
using the time-ordered graviton propagator discussed in~\citep{Marshman:2019sne}.
The Einstein-Hilbert action can be written as in terms of the fluctuations
$h_{\mu\nu}$ up to quadratic in order: 
\[
S=(1/4)\int d^{4}x~h_{\mu\nu}{\cal O}^{\mu\nu\lambda\sigma}h_{\lambda\sigma}+{\cal O}(h^{3})
\]
\\
 where ${\cal O}^{\mu\nu\lambda\sigma}=(1/4)(\eta^{\mu\rho}\eta^{\nu\sigma}+\eta^{\mu\sigma}\eta^{\nu\rho})\Box-(1/2)\eta^{\mu\nu}\eta^{\rho\sigma}\Box+(1/2)(\eta^{\mu\nu}\partial^{\rho}\partial^{\sigma}+\eta^{\rho\sigma}\partial^{\mu}\partial^{\nu}-\eta^{\mu\rho}\partial^{\nu}\partial^{\sigma}-\eta^{\mu\sigma}\partial^{\nu}\partial^{\rho})$,
\\
 where the d'Alembertain operator is: $\Box=g_{\mu\nu}\nabla^{\mu}\nabla^{\nu}$.
The propagator for the graviton~\citep{Scadron,Marshman:2019sne}
$h_{\mu\nu}$ can be recast in terms of 
\[
\Pi{\mu\nu\rho\sigma}({\bm{k}})=(1/2{\bm{k}}^{2})(\eta_{\mu\rho}\eta_{\nu\sigma}+\eta_{\nu\rho}\eta_{\mu\sigma}-\eta_{\mu\nu}\eta_{\rho\sigma}).
\]
With the help of this propagator, one can find the gravitational potential,
i.e. the non-relativistic scattering due to an exchange of an off-shell
graviton. The gravitational potential is given by $\Phi({\bm{r}})=-(8\pi G/(2\pi)^{3})\int d^{3}{\bm{k}}T_{1}^{00}\Pi_{0000}(k)T_{2}^{00}(-k)e^{i{\bm{k}}\cdot{\bm{r}}}=-4\pi Gm^{2}\int d^{3}{\bm{k}}e^{i{\bm{k}}\cdot{\bm{r}}}/{\bm{k}^{2}}=-Gm^{2}/{\bm{r}}$.
This result is the same as what we have obtained in Eq.~(\ref{eq:Newton}).
The only difference here is that we have computed the potential by
using the full graviton propagator and the scattering amplitude between
the two masses via the exchange of a spin-2 and spin-0 components
of the graviton, see the appendix of Ref.~\cite{Marshman:2019sne}.
In the text we have computed the change in the graviton vacuum. However,
in the non-relativistic limit both the results give rise to the same
conclusion.}: 
\begin{equation}
\Delta{\hat{H}_{g}}=-\frac{Gm^{2}}{\vert{\hat{x}}_{A}-{\hat{x}}_{B}\vert}.\label{eq:Newton}
\end{equation}
We thus find that the change in the graviton energy, $\Delta\hat{H}_{g}$,
due to the interaction between the graviton and the matter is an operator
valued function of the two matter systems, i.e. 
\begin{equation}
\Delta\hat{H}_{g}\equiv f({\hat{x}}_{A},{\hat{x}}_{B}).
\end{equation}
If the two matter systems do not have a sharply defined positions
(such as when placed in a spatial superposition or some other non-classical
state) then the change in the graviton energy $\Delta\hat{H}_{g}$
will not be a real number, as required in a classical theory of gravity,
but rather an operator-valued quantity, a bonafide quantum entity.

We now wish to calculate the excited wave function $|\psi_{f}\rangle$
of the two harmonic oscillators to establish the link between entanglement
and LOQC discussed in Sec.~\ref{sec:Interaction-induces-entanglement}.
We first use Eq.~(\ref{eq:xy}) and expand Eq.~(\ref{eq:Newton})
to find 
\begin{equation}
\Delta\hat{H}_{g}\approx-\frac{Gm^{2}}{d}+\frac{Gm^{2}}{d^{2}}(\delta\hat{x}_{B}-\delta\hat{x}_{A})-\frac{Gm^{2}}{d^{3}}(\delta\hat{x}_{B}-\delta\hat{x}_{A})^{2}.
\end{equation}
The last term gives the lowest-order matter-matter interaction~\footnote{It is instructive to compare the obtained results for two harmonic
oscillators to the results obtained previously for two interferometers.
In both cases, the action is proportional to $S=E\tau/\hbar$, where
the interaction energy of the system is given by $E\sim H_{AB}$ and
$\tau$ is the coherence time scale. Considering the setup in~\citep{Bose:2017nin,Marletto},
and setting $\Delta\phi\sim S$, we then recover the entanglement
phase $\Delta\phi\sim(2Gm^{2}/\hbar d)(\delta x/d)^{2}\tau$, where
we have assumed $\delta x_{A}\sim\delta x_{B}\sim\delta x$ for the
localised spatial superpositions of the two test masses. \label{fn6}} 
\begin{equation}
\hat{H}_{\text{AB}}\equiv\frac{2Gm^{2}}{d^{3}}\delta\hat{x}_{A}\delta\hat{x}_{B}.\label{eq:HabN}
\end{equation}
Note that the interaction Hamiltonian $\hat{H}_{\text{AB}}$ contains
only the operators of the two harmonic oscillators $\delta\hat{x}_{A},~\delta\hat{x}_{B}$.
Yet it is critical to realise that the product $\delta\hat{x}_{A}\delta\hat{x}_{B}$
would not have arisen if we had assumed a real-valued shift of the
energy of the gravitational field. Indeed, a classical gravitational
field is unable to produce the operator-valued shift in Eq.~\eqref{eq:Newton}
(and hence the quantum interaction potential in Eq.~\eqref{eq:HabN}).
We must thus conclude that gravitationally induced entanglement is
indeed a quantum signature of the gravitational field~\footnote{The above expression, Eq.~(\ref{eq:HabN}), has been the starting
point for the entanglement of the two harmonic oscillators with $1/r$-potential
in many analyses, see~\cite{Kafri:2014zsa,Qvarfort,Dutta,Krishnanda,Cosco,Weiss},
but here we have shown how this interaction arises by noting that
how the vacuum of the spin-2 and spin-0 components of the graviton
has shifted due to the quantum nature of the harmonic oscillators. }.

We will now use the modes in Eq.~(\ref{eq:modesxy}) to find 
\begin{equation}
\hat{H}_{\text{AB}}\approx\hbar\mathfrak{g}(\hat{a}\hat{b}+\hat{a}^{\dagger}\hat{b}+\hat{a}\hat{b}^{\dagger}+\hat{a}^{\dagger}\hat{b}^{\dagger}),
\end{equation}
where we have defined the coupling 
\begin{equation}
\mathfrak{g}\equiv\frac{Gm}{d^{3}\omega_{\text{m}}}.\label{eq:couplingNewton}
\end{equation}
Using $\hat{H}_{\text{AB}}$ as the interaction Hamiltonian in Eq.~(\ref{eq:coefficients})
we find that the only non-zero coefficient emerges from the term $\sim\hat{a}^{\dagger}\hat{b}^{\dagger}$
and is given by: 
\begin{alignat}{1}
C_{11} & =-\frac{\mathfrak{g}}{2\omega_{\text{m}}}.
\end{alignat}
We note that the $a^{\dagger}b^{\dagger}$ term generates the first
excited states in the harmonic oscillators (with energy $E_{1}=E_{0}+\hbar\omega_{\text{m}}$).
In addition, we also have the term $C_{00}=1$ corresponding to the
unperturbed state.

The final state in Eq.~(\ref{eq:final}) thus simplifies to (up to
first order in the perturbation theory, and by setting $\lambda=1)$:
\begin{equation}
\vert\psi_{\text{f}}\rangle\equiv\frac{1}{\sqrt{1+(\mathfrak{g}/(2\omega_{\text{m}}))^{2}}}[\vert0\rangle\vert0\rangle-\frac{\mathfrak{g}}{2\omega_{\text{m}}}\vert1\rangle\vert1\rangle],\label{eq:finalNewton}
\end{equation}
which is \emph{an entangled state involving the ground and the first
excited states of the two harmonic oscillators}. We compute the reduced
density matrix by tracing system B (we recall that our notation is
$\vert n\rangle\vert N\rangle=\vert n\rangle_{A}\vert N\rangle_{B}$).
The concurrence in Eq.~(\ref{eq:concurrence2}) reduces to 
\begin{equation}
C\equiv\sqrt{2(1-\frac{1+(\mathfrak{g}/(2\omega_{\text{m}}))^{4}}{1+(\mathfrak{g}/(2\omega_{\text{m}}))^{2}})}\approx\sqrt{2}\frac{\mathfrak{g}}{\omega_{\text{m}}},\label{eq:concurrenceNewton}
\end{equation}
which is valid when the parameter ${\mathfrak{g}}/{\omega_{\text{m}}}\ll1$
is small. Inserting the coupling from Eq.~(\ref{eq:couplingNewton})
we find the concurrence is given by: 
\begin{equation}
C=\frac{\sqrt{2}Gm}{d^{3}\omega_{\text{m}}^{2}}.\label{eq:CN}
\end{equation}
We thus see that the the degree of entanglement grows linearly with
the mass of the oscillator and inversely with the distance between
the two oscillators (inverse cubic) as well as with the frequency
of the harmonic trap (inverse square).

Let us reiterate the key finding. If the underlying gravitational
field were classical (specifically, obeying LOCC), then the final
state of the matter components, i.e. the two harmonic oscillator states,
would have never evolved to the entangled state $|\psi_{f}\rangle$,
but would have rather remained in an unentangled/separable state.
Conversely, if the gravitational field is quantized (and hence obeys
LOQC) then we have shown that it can give rise to the entangled state
$|\psi_{f}\rangle$. 

\section{Entanglement via graviton in the non-static case\label{subsec:Entanglement-from-hij}}

In this section, we are interested in the coupling of the gravitational
field to the $\hat{T}_{ij}$ components of the stress energy tensor.
In our specific case we consider two particles (in harmonic traps)
moving along the $x$-axis such that the only non-zero components
are given by $T_{00}$, $T_{01}$ and $T_{11}$ (with $T_{10}=T_{01}$).
Hence the relevant components of the graviton are given by $\hat{h}_{00}=\hat{\gamma}_{00}+(1/2)\hat{\gamma}$
(already present in the static case), by $\hat{h}_{01}=\hat{h}_{10}=\hat{\gamma}_{01}$,
and by $\hat{h}_{11}=\hat{\gamma}_{11}-(1/2)\hat{\gamma}$ (which
can be identified with the degrees of freedom of the GWs as discussed
below). We will find that the energy shift in the graviton vacuum
induces a coupling between the two harmonic oscillator states, which
leads to the entanglement only when we assume that the $\hat{h}_{00},\hat{h}_{01},\hat{h}_{11}$
components are quantum.

The computation follows the analogous steps as the ones discussed
in the previous section. 
The basic assumption is that these graviton modes are quantized, and
act as a quantum communicator, or serve as a quantum interaction between
the two harmonic oscillators. The interaction Hamiltonian has now
two contributions: 
\begin{alignat}{1}
\hat{H}_{\text{int}}= & \frac{1}{2}\int d{\bm{r}}\left[\hat{\gamma}_{00}({\bm{r}})+(1/2)\hat{\gamma}({\bm{r}})\right]\hat{T}_{00}({\bm{r}})\nonumber \\
 & +\int d{\bm{r}}\hat{\gamma}_{01}({\bm{r}})\hat{T}_{01}({\bm{r}})\nonumber \\
 & +\frac{1}{2}\int d{\bm{r}}\left[\hat{\gamma}_{11}({\bm{r}})-(1/2)\hat{\gamma}({\bm{r}})\right]\hat{T}_{11}({\bm{r}}),\label{eq:Hint2}
\end{alignat}
where the first line coincides with the interaction considered in
Eq.~\eqref{eq:H00}, while the second and third lines arise from
the degrees of freedom of the GWs corresponding to the + polarization~\footnote{We recall that in the the TT gauge we have the interaction Hamiltonian
given by\cite{Weinberg} 
\begin{equation}
\hat{H}_{\text{int}}=-\frac{1}{2}\int d\bm{r}\hat{h}_{ij}(\bm{r})\hat{T}_{ij}(\bm{r}),\label{eq:Hij}
\end{equation}
where we implicitly assume the summation over the indices $i,j=1,2,3$.
The propagating, on-shell, graviton is described by the two helicity
states $(+,\times)$: 
\begin{equation}
\hat{h}_{ij}={\cal A}\int d{\bm{k}}\sqrt{\frac{\hbar}{2\omega_{\bm{k}}(2\pi)^{3}}}P_{\lambda}^{\dagger}(\bm{k})e_{ij}^{\lambda}(\bm{k})e^{-i{\bm{k}}\cdot{\bm{r}}}+{\rm H.c},\label{eq:hij}
\end{equation}
where we have assumed the summation over the two polarizations $(+,~\times)$
($e_{jk}^{\lambda}$ denote the basis for the two polarization states),
and the annihilation and the creation operator satisfies 
\begin{equation}
[\hat{P}_{\lambda}(\bm{k}),\hat{P}_{\lambda}^{\dagger}(\bm{k}')]=\delta(\bm{k}-\bm{k}').
\end{equation}
The trace-reversed perturbation $\hat{h}_{ij}(\bm{r})$ in Eq.~\eqref{eq:hij}
can be identified with $\hat{\gamma}_{ij}({\bm{r}})-(1/2)\eta_{ij}\hat{\gamma}({\bm{r}})$.
In particular, in our specific case the + polarization GW $\hat{h}_{11}(\bm{r})$
can be identified with $\hat{\gamma}_{11}({\bm{r}})-(1/2)\hat{\gamma}({\bm{r}})$.\label{fn:GWs}}.

Let us first rewrite the interaction term in Eq.~\eqref{eq:Hint2}
using the definitions in Eqs.~(\ref{spin-2}) and (\ref{spin-0}):
\begin{alignat}{1}
\hat{H}_{\text{int}}= & \frac{\mathcal{A}}{2}\int d\bm{k}\sqrt{\frac{\hbar}{2\omega_{\bm{k}}}}(\left[\hat{P}_{00}(\bm{k})+\hat{P}(\bm{k})\right]\hat{T}_{00}(\bm{k})+\text{H.c})\nonumber \\
 & +\mathcal{A}\int d\bm{k}\sqrt{\frac{\hbar}{2\omega_{\bm{k}}}}(\hat{P}_{01}(\bm{k})\hat{T}_{01}(\bm{k})+\text{H.c})\nonumber \\
 & +\frac{\mathcal{A}}{2}\int d\bm{k}\sqrt{\frac{\hbar}{2\omega_{\bm{k}}}}(\left[\hat{P}_{11}(\bm{k})-\hat{P}(\bm{k})\right]\hat{T}_{11}(\bm{k})+\text{H.c}),\label{eq:Hamiltonian}
\end{alignat}
where we have introduced the Fourier transform of the stress-energy
tensor 
\begin{equation}
\hat{T}_{\mu\nu}(\bm{k})=\frac{1}{\sqrt{(2\pi)^{3}}}\int d\bm{r}e^{-i\bm{k}\cdot\bm{r}}\hat{T}_{\mu\nu}(\bm{r}).\label{eq:Fourier2}
\end{equation}
Since we are considering the two harmonic oscillators to be moving
along the $x$-axis such that the only non-zero components are given
by 
\begin{equation}
\hat{T}_{\mu\nu}(\bm{r})\equiv\frac{\hat{p}_{\mu}\hat{p}_{\nu}}{E/c^{2}}(\delta(\bm{r}-\hat{\bm{r}}_{A})+\delta(\bm{r}-\hat{\bm{r}}_{B})),\label{eq:currentR}
\end{equation}
where $p_{\mu}=(-E/c,\bm{p})$, $E=\sqrt{\bm{p}^{2}c^{2}+m^{2}c^{4}}$,
$\mu,\nu=0,1$, and $\hat{\bm{r}}_{A}=(\hat{x}_{A},0,0)$,~$\hat{\bm{r}}_{B}=(\hat{x}_{B},0,0)$
denote the positions of the two matter systems. Here we have promoted
the classical expression of the stress-energy tensor to a quantum
operator following the Weyl quantization prescription to ensure that
the quantum stress-energy tensor is a Hermitian operator. In order
to simplify the notation we will however write the unsymmetrized expressions
(e.g. $\hat{x}\hat{p}$), implicitly assuming that all expressions
need to be interpreted in the symmetrized ordering (e.g. $(\hat{x}\hat{p}+\hat{p}\hat{x})/2$).
Using Eqs.~\eqref{eq:Fourier2} and \eqref{eq:currentR}, we find
the following Fourier space components: 
\begin{alignat}{1}
\hat{T}_{00}(\bm{k}) & =\frac{1}{\sqrt{(2\pi)^{3}}}(\hat{E}_{A}e^{i\bm{k}\cdot\hat{\bm{r}}_{A}}+\hat{E}_{B}e^{i\bm{k}\cdot\hat{\bm{r}}_{B}}),\label{eq:sources1}\\
\hat{T}_{01}(\bm{k}) & =-\frac{1}{\sqrt{(2\pi)^{3}}}(\hat{p}_{A}ce^{i\bm{k}\cdot\hat{\bm{r}}_{A}}+\hat{p}_{B}ce^{i\bm{k}\cdot\hat{\bm{r}}_{B}}),\\
\hat{T}_{11}(\bm{k}) & =\frac{1}{\sqrt{(2\pi)^{3}}}(\frac{\hat{p}_{A}^{2}c^{2}}{E_{A}}e^{i\bm{k}\cdot\hat{\bm{r}}_{A}}+\frac{\hat{p}_{B}^{2}c^{2}}{E_{B}}e^{i\bm{k}\cdot\hat{\bm{r}}_{B}}).\label{eq:sources3}
\end{alignat}
We can readily extend the computation from Sec.~\ref{subsec:Entanglement-from-h00}
to Eq.~\eqref{eq:Hamiltonian} by including in the computation the
intermediate graviton states: $\vert\bm{k}\rangle=\frac{1}{\sqrt{2}}\hat{P}_{00}^{\dagger}(\bm{k})\vert0\rangle$,
$\frac{1}{\sqrt{2}}\hat{P}_{11}^{\dagger}(\bm{k})\vert0\rangle$,
$\hat{P}_{01}^{\dagger}(\bm{k})\vert0\rangle$, and $\hat{P}^{\dagger}(\bm{k}))\vert0\rangle$
(where the prefactor $\frac{1}{\sqrt{2}}$ in the first two states
ensured the correct normalization~\footnote{The normalization of the states can be computed using the commutation
relations in Eqs.\eqref{eq:comm1} and \eqref{eq:comm2}. Let us consdier
first $\hat{P}_{00}^{\dagger}(\bm{k})\vert0\rangle$. We note that
$\langle0\vert\hat{P}_{00}(\bm{k})\hat{P}_{00}^{\dagger}(\bm{k}')\vert0\rangle=\langle0\vert[\hat{P}_{00}(\bm{k}),\hat{P}_{00}^{\dagger}(\bm{k}')]\vert0\rangle$,
where we have used the definition of the commutator $[\hat{O}_{1},\hat{O}_{2}]=\hat{O}_{1}\hat{O}_{2}-\hat{O}_{2}\hat{O}_{1}$
(as well as the definition of the of the vacuum state). Using \eqref{eq:comm1}
we then readily find $\langle0\vert\hat{P}_{00}(\bm{k})\hat{P}_{00}^{\dagger}(\bm{k}')\vert0\rangle=2\delta^{(3)}(\bm{k}-\bm{k}')$.
Using analogous steps we find $\langle0\vert\hat{P}_{11}(\bm{k})\hat{P}_{11}^{\dagger}(\bm{k}')\vert0\rangle=2\delta^{(3)}(\bm{k}-\bm{k}')$,
$\langle0\vert\hat{P}_{01}(\bm{k})\hat{P}_{01}^{\dagger}(\bm{k}')\vert0\rangle=-\delta^{(3)}(\bm{k}-\bm{k}')$,
and $\langle0\vert\hat{P}(\bm{k})\hat{P}^{\dagger}(\bm{k}')\vert0\rangle=-\delta^{(3)}(\bm{k}-\bm{k}')$.\label{fn:normalizations}}). The energy-shift of the graviton vacuum $\vert0\rangle$ is thus
given by the second-order perturbation theory (while the first order
perturbation will vanish \footref{fn3}): 
\begin{equation}
\Delta\hat{H}_{g}\equiv\sum\int d\bm{k}\frac{\langle0\vert\hat{H}_{\text{int}}\vert\bm{k}\rangle\langle\bm{k}\vert\hat{H}_{\text{int}}\vert\bm{0}\rangle}{E_{0}-E_{\bm{k}}},\label{eq:dHij}
\end{equation}
where the sum indicates summation over the one particle projectors
\footnote{The projectors $\frac{\vert\bm{k}\rangle\langle\bm{k}\vert}{\langle\bm{k}\vert\bm{k}\rangle}$
are given by $\frac{1}{2}\hat{P}_{00}^{\dagger}(\bm{k})\vert0\rangle\langle0\vert\hat{P}_{00}(\bm{k})\vert$,
$\frac{1}{2}\hat{P}_{11}^{\dagger}(\bm{k})\vert0\rangle\langle0\vert\hat{P}_{11}(\bm{k})\vert$,
$-\hat{P}_{01}^{\dagger}(\bm{k})\vert0\rangle\langle0\vert\hat{P}_{01}(\bm{k})\vert$
and $-\hat{P}^{\dagger}(\bm{k})\vert0\rangle\langle0\vert\hat{P}(\bm{k})\vert$.
The normalization prefactors $\frac{1}{\langle\bm{k}\vert\bm{k}\rangle}=\frac{1}{2},$$\frac{1}{2},-1,-1$
are a direct consequence of the commutation relations in Eqs.~\eqref{eq:comm1}
and \eqref{eq:comm2} which fix the normalization of the states\footref{fn:normalizations}.
With this definitions of the projectors we find that $\frac{\vert\bm{k}\rangle\langle\bm{k}\vert}{\langle\bm{k}\vert\bm{k}\rangle}\vert\bm{k}\rangle=+1\vert\bm{k}\rangle$,
i.e. the projectors give a positive eigenvalue $+1$ as expected.
In Eq.~\eqref{eq:dHij} we are thus implicitly using the normalized
projectors $\frac{\vert\bm{k}\rangle\langle\bm{k}\vert}{\langle\bm{k}\vert\bm{k}\rangle}$
when we write $\vert\bm{k}\rangle\langle\bm{k}\vert$.} $\vert\bm{k}\rangle\langle\bm{k}\vert$ constructed on the unperturbed
vacuum, $E_{0}$ is the energy of the vacuum state, and $E_{\bm{k}}=E_{0}+\hbar\omega_{\bm{k}}$
is the energy of the one-particle state. We can readily evaluate 
\begin{alignat}{1}
\langle0\vert\hat{P}(\bm{k})\hat{H}_{\text{int}}\vert\bm{0}\rangle & =\frac{\mathcal{A}}{2}\sqrt{\frac{\hbar}{2\omega_{\bm{k}}}}(\hat{T}_{00}(\bm{k})-\hat{T}_{11}(\bm{k})),\label{eq:expectation1}\\
\langle0\vert\hat{P}_{01}(\bm{k})\hat{H}_{\text{int}}\vert\bm{0}\rangle & =\mathcal{A}\sqrt{\frac{\hbar}{2\omega_{\bm{k}}}}\hat{T}_{01}(\bm{k}),\\
\langle0\vert\hat{P}_{00}(\bm{k})\hat{H}_{\text{int}}\vert\bm{0}\rangle & =\mathcal{A}\sqrt{\frac{\hbar}{2\omega_{\bm{k}}}}\hat{T}_{00}(\bm{k}),\\
\langle0\vert\hat{P}_{11}(\bm{k})\hat{H}_{\text{int}}\vert\bm{0}\rangle & =\mathcal{A}\sqrt{\frac{\hbar}{2\omega_{\bm{k}}}}\hat{T}_{11}(\bm{k}).\label{eq:expectation3}
\end{alignat}
By using Eqs.~\eqref{eq:expectation1}-\eqref{eq:expectation3},
we then find from Eq.~\eqref{eq:dHij}: 
\begin{alignat}{1}
\Delta\hat{H}_{g}= & -\mathcal{A}^{2}\int d\bm{k}\frac{\hat{T}_{00}^{\dagger}(\bm{k})\hat{T}_{00}(\bm{k})+\hat{T}_{11}^{\dagger}(\bm{k})\hat{T}_{11}(\bm{k})}{8c^{2}\bm{k}^{2}}\nonumber \\
 & -\mathcal{A}^{2}\int d\bm{k}\frac{(\hat{T}_{00}^{\dagger}(\bm{k})\hat{T}_{11}(\bm{k})+\text{H.c.})}{8c^{2}\bm{k}^{2}}\nonumber \\
 & +4\mathcal{A}^{2}\int d\bm{k}\frac{\hat{T}_{01}^{\dagger}(\bm{k})\hat{T}_{01}(\bm{k})}{8c^{2}\bm{k}^{2}}.\label{eq:dH3}
\end{alignat}
We now use the fact that the two particles are confined along the
x-axis, where we set ${\hat{p}}_{Ay}={\hat{p}}_{Az}={\hat{p}}_{By}={\hat{p}}_{Bz}=0$,
and write ${\hat{p}}_{A}\equiv{\hat{p}}_{Ax},{\hat{p}}_{B}\equiv{\hat{p}}_{Bx}$
, ${\hat{\bm{r}}}_{A}=({x}_{A},0,0)$,${\hat{\bm{r}}}_{B}=({x}_{B},0,0)$
and, $\bm{k}=(k_{x},k_{y},k_{z})$. We then insert Eqs.~\eqref{eq:sources1}-\eqref{eq:sources3}
to find \footnote{In Eq.~\eqref{eq:56} we have omitted cross-terms between each particle
with itself, i.e., terms involving only one of the two particles such
as $\sim\hat{E}_{A}\hat{E}_{A},\hat{E}_{B}\hat{E}_{B},\dots$ Such
terms are known as the self-energy terms and do not contribute to
the interaction between the two particles. Analogous self-energy terms
appear also in electromagnetism when we try to compute the interaction
between the two charges (see for example Ref. \cite{Cohen}). } 
\begin{alignat}{1}
\Delta\hat{H}_{g}= & -\frac{\mathcal{A}^{2}}{(2\pi)^{3}}\int d\bm{k}\left(\frac{\hat{E}_{A}\hat{E}_{B}+\frac{\hat{p}_{A}^{2}c^{2}}{E_{A}}\frac{\hat{p}_{B}^{2}c^{2}}{E_{B}}}{8c^{2}\bm{k}^{2}}\right.\nonumber \\
 & \left.+\frac{\hat{E}_{A}\frac{\hat{p}_{B}^{2}c^{2}}{E_{B}}+\hat{E}_{B}\frac{\hat{p}_{A}^{2}c^{2}}{E_{A}}}{8c^{2}\bm{k}^{2}}-4\frac{\hat{p}_{A}c\hat{p}_{B}c}{8c^{2}\bm{k}^{2}}\right)\nonumber \\
 & (e^{ik_{x}(\hat{x}_{A}-\hat{x}_{B})}+e^{-ik_{x}(\hat{x}_{A}-\hat{x}_{B})}).\label{eq:56}
\end{alignat}
Performing the integration and expanding in powers of $1/c^{2}$,
we find that Eq.~(\ref{eq:56}) simplifies to \footnote{It is instructive to compare the gravitational potential obtained
in Eq.~\eqref{eq:dH5} to the results for classical point particles
in the literature. We first transfrom from the reference frame of
the two traps to the center-of-mass reference frame where we have
$p\equiv p_{A}=-p_{B}$, and denote $r\equiv\vert{x}_{A}-{x}_{B}\vert$.
From Eq.~\eqref{eq:dH5} we then find the potential 
\begin{equation}
\Delta H_{g}=-\frac{Gm^{2}}{r}-7\frac{Gp^{2}}{c^{2}r}-\frac{Gp^{4}}{c^{4}m^{2}r},\label{eq:dHgC}
\end{equation}
which matches the results previously obtained using different methods~\cite{Iwasaki1971,Hiida1972,Cristofoli2019,Grignani2020}. }: 
\begin{alignat}{1}
\Delta\hat{H}_{g} & =-\frac{Gm^{2}}{\vert{\hat{x}}_{A}-{\hat{x}}_{B}\vert}\nonumber \\
 & -\frac{G(3{\hat{p}}_{A}^{2}-8\hat{p}_{A}\hat{p}_{B}+3{\hat{p}}_{B}^{2})}{2c^{2}\vert{\hat{x}}_{A}-{\hat{x}}_{B}\vert}\nonumber \\
 & -\frac{G(5{\hat{p}}_{A}^{4}-18{\hat{p}}_{A}^{2}{\hat{p}}_{B}^{2}+5{\hat{p}}_{B}^{4})}{8c^{4}m^{2}\vert{\hat{x}}_{A}-{\hat{x}}_{B}\vert}.\label{eq:dH5}
\end{alignat}
Eq.~\eqref{eq:56} contains the exact couplings between the two masses
up to order $\mathcal{O}(1/c^{4})$ and to the leading order IR contributions
in Newton's constant, $G$. Note that if we set ${\hat{p}}_{A}={\hat{p}}_{B}=0$,
the last two terms vanish. However, a quantum system retains its zero
point fluctuations and hence we find $\langle{\hat{p}}_{A}^{2}\rangle=\langle{\hat{p}}_{B}^{2}\rangle\sim\hbar m\omega_{m}$
even for ground states of the two harmonic oscillators (using Eq.~\eqref{eq:modepxpy}
and the canonical commutation relations).

Let us make a brief comment on Eq.~(\ref{eq:dH5}). By quantising
the graviton we have obtained 
\begin{equation}
\Delta\hat{H}_{g}\equiv f({\hat{p}}_{A},{\hat{p}}_{B},{\hat{x}}_{A},{\hat{x}}_{B}),\label{eq:dHGW}
\end{equation}
which is an operator-valued shift in the vacuum energy depending on
the matter operators. On the other hand, if we would have assumed
a classical gravitational field we could have only generated a real-valued
shift $\Delta H_{g}$ in a complete analogy to we have discussed in
Eq.~\eqref{eq:Newton}.

We will be interested in computing the lowest order corrections for
the final matter state $\vert\psi_{f}\rangle$ due to the second and
third term on the right-hand side of Eq.~\eqref{eq:dH5} (the first
term has been already discussed in Sec.~\ref{subsec:Entanglement-from-h00}).

\section{Computing the concurrence for case-1 \label{case1}}

We first discuss the second term on the right-hand side of Eq.~\eqref{eq:dH5}.
We can extract the lowest order non-trivial quantum interaction term~\footnote{Intuitively, it is again interesting to estimate the entanglement
phase. We find $\Delta\phi\sim4Gp_{a}p_{B}\tau/(c^{2}\hbar d)$, where
$\tau$ is the coherence time scale. As expected such effects are
thus typically suppressed in comparison to the phase accumulated from
the exchange of graviton in the static case \footref{fn6}.}: 
\begin{equation}
\hat{H}_{AB}\sim4\frac{G\hat{p}_{A}\hat{p}_{B}}{c^{2}d}+\cdots.\label{eq:Hab}
\end{equation}
Note that at the lowest order in the expansion of the denominator,
$\hat{x}_{A},~\hat{x}_{B}$ do not occur, and the interaction Hamiltonian
is dominated by the momentum operators $\hat{p}_{A},~\hat{p}_{B}$.
We will now use the modes in Eq.~(\ref{eq:modepxpy}) to find 
\begin{equation}
\hat{H}_{AB}\approx\hbar\mathfrak{g}(\hat{a}^{\dagger}-\hat{a})(\hat{b}^{\dagger}-\hat{b}),\label{eq:Hab2}
\end{equation}
where the coupling is given by 
\begin{equation}
\mathfrak{g}=\frac{2Gm\omega_{\text{m}}}{c^{2}d}.\label{eq:couplingGW}
\end{equation}
As we will see the only term that is relevant in our case is $\hat{a}^{\dagger}\hat{b}^{\dagger}$,
which signifies that the final matter state is a linear combination
of $|0\rangle|0\rangle$ and $|1\rangle|1\rangle$. In particular,
using $\hat{H}_{\text{AB}}$ as the interaction Hamiltonian in Eq.~(\ref{eq:coefficients})
we find that the only non-zero perturbation coefficient emerges from
the term $\sim\hat{a}^{\dagger}\hat{b}^{\dagger}$, and it is given
by: 
\begin{alignat}{1}
C_{11} & =-\frac{\mathfrak{g}}{2\omega_{\text{m}}}.
\end{alignat}
Here we have used the fact that energy momentum conservation constraints:
\begin{equation}
E_{1}=E_{0}+\hbar\omega_{\text{m}}\,.
\end{equation}
Note that it is twice the frequency of the harmonic oscillators. In
addition, we also have the term $C_{00}=1$, corresponding to the
unperturbed state.

We find that the final state in Eq.~(\ref{eq:final}) thus simplifies
to (setting $\lambda=1$): 
\begin{equation}
\vert\psi_{\text{f}}\rangle\equiv\frac{1}{\sqrt{1+(\mathfrak{g}/(2\omega_{\text{m}}))^{2}}}[\vert0\rangle\vert0\rangle-\frac{\mathfrak{g}}{2\omega_{\text{m}}}\vert1\rangle\vert1\rangle],\label{eq:finalMixed}
\end{equation}
which is \emph{an entangled state involving the ground and the first
excited states of the harmonic oscillators} (up to first order in
the perturbation theory). We compute the reduced density matrix by
tracing away system B (we recall that our notation is $\vert n\rangle\vert N\rangle=\vert n\rangle_{A}\vert N\rangle_{B}$).
The concurrence in Eq.~(\ref{eq:concurrence2}) reduces to 
\begin{equation}
C\equiv\sqrt{2(1-\frac{1+(\mathfrak{g}/(2\omega_{\text{m}}))^{4}}{1+(\mathfrak{g}/(2\omega_{\text{m}}))^{2}})}\approx\sqrt{2}\frac{\mathfrak{g}}{\omega_{\text{m}}},\label{eq:concurrenceNewton-1-1}
\end{equation}
which is valid when the parameter ${\mathfrak{g}}/{\omega_{\text{m}}}\ll1$.
After inserting the coupling from Eq.~(\ref{eq:couplingGW}), we
find the concurrence to be: 
\begin{equation}
C=\frac{2\sqrt{2}Gm}{c^{2}d}.
\end{equation}
Note that the the degree of entanglement grows linearly with the mass
of the harmonic oscillators, does not depend on the frequency, and
scales inversely with the distance between the two oscillators.

We find that the concurrence in the case of a static limit given in
Eq.~(\ref{eq:CN}) dominates over the non-static case, provided 
\begin{equation}
\frac{\omega_{{\rm m}}d}{c}<\frac{1}{\sqrt{2}}.\label{threshold2}
\end{equation}
For example, with $\omega_{{\rm m}}\sim10^{8}\,\text{Hz}$ we find that
the threshold value is obtained already at $d\sim1\,\text{m}$. Hence,
such effects could in principle be tested already with a small tabletop
experiment, but the feasibility of the experiment has to be studied
separately.


\section{Computing the concurrence for case-2 \label{case2}}

From the last term in Eq.~\eqref{eq:dH5} we can extract the lowest
order non-trivial quantum interaction term\footnote{It is again interesting to estimate the approximate entanglement phase.
We find $\Delta\phi\sim9Gp_{a}^{2}p_{B}^{2}\tau/(4c^{4}m^{2}\hbar d)$,
where $\tau$ is the coherence time scale. As expected such effects
are thus typically suppressed in comparison to the phase accumulated
from the exchange of the graviton in the static case \footref{fn6}.}: 
\begin{equation}
\hat{H}_{AB}\sim-\frac{9G\hat{p}_{A}^{2}\hat{p}_{B}^{2}}{4c^{4}m^{2}d}+\cdots.\label{eq:Hab-2}
\end{equation}
Note that at the lowest order in the expansion of the denominator,
$\hat{x}_{A},~\hat{x}_{B}$ do not occur, and the interaction Hamiltonian
is dominated by the momentum operators $\hat{p}_{A},~\hat{p}_{B}$.
We will now use the modes in Eq.~(\ref{eq:modepxpy}) to find 
\begin{equation}
\hat{H}_{AB}\approx-\hbar\mathfrak{g}(\hat{a}^{\dagger}-\hat{a})^{2}(\hat{b}^{\dagger}-\hat{b})^{2},\label{eq:Hab2222}
\end{equation}
where the coupling is given by 
\begin{equation}
\mathfrak{g}=\frac{9G\hbar\omega_{\text{m}}^{2}}{16c^{4}d}.\label{eq:couplingGW-2}
\end{equation}
The only term that is relevant in our case is $(\hat{a}^{\dagger}\hat{b}^{\dagger})^{2}$,
which signifies that the final matter state is a linear combination
of $|0\rangle|0\rangle$ and $|2\rangle|2\rangle$. Hence, at the
lowest order the gravitons carry twice the energy of the harmonic
oscillators, i.e. $\omega_{\bm{k}}=2\omega_{\bm{m}}$.

In particular, using $\hat{H}_{\text{AB}}$ as the interaction Hamiltonian
in Eq.~(\ref{eq:coefficients}), we find that the only non-zero perturbation
coefficient emerges from the term $\sim\hat{a}^{\dagger2}\hat{b}^{\dagger2}$
and is given by: 
\begin{alignat}{1}
C_{22} & =\frac{\mathfrak{g}}{2\omega_{\text{m}}}.
\end{alignat}
Here we have used the fact that energy momentum conservation constraints:
\begin{equation}
E_{2}=E_{0}+2\hbar\omega_{\text{m}}\,.
\end{equation}
Note that it is twice the frequency of the harmonic oscillators. In
addition, we also have the term $C_{00}=1$, corresponding to the
unperturbed state.

We find that the final state in Eq.~(\ref{eq:final}) thus simplifies
to (setting $\lambda=1$): 
\begin{equation}
\vert\psi_{\text{f}}\rangle\equiv\frac{1}{\sqrt{1+(\mathfrak{g}/(2\omega_{\text{m}}))^{2}}}[\vert0\rangle\vert0\rangle+\frac{\mathfrak{g}}{2\omega_{\text{m}}}\vert2\rangle\vert2\rangle],\label{eq:finalGW}
\end{equation}
which is \emph{an entangled state involving the ground and the second
excited states of the harmonic oscillators} (up to first order in
the perturbation theory).

Note that the occurrence of the second excited states from the initial
ground states requires the transition $n\rightarrow n+2$, where $n$
is the number eigenvalue of the harmonic oscillator. This distinct
$n\rightarrow n+2$ transition can be traced back to the coupling
to the gravitational field, see Eqs.~(\ref{eq:dH5}), (\ref{eq:Hab-2}),
and (\ref{eq:Hab2222}). In particular, it emerges from the coupling
$\propto\hat{h}_{11}\hat{T}_{11}$, where $\hat{h}_{11}$ can be identified
with the degrees of freedom associated to the ``+'' gravitational
waves~\footref{fn:GWs}. In our case we have $\hat{T}_{11}\sim(\hat{a}^{\dagger})^{2},(\hat{b}^{\dagger})^{2}$,
and thus we find the couplings $(\hat{a}^{\dagger})^{2}\hat{h}_{11}$
and $(\hat{b}^{\dagger})^{2}\hat{h}_{11}$, which lead to the transition
$n\rightarrow n+2$ for the two harmonic oscillators. In general, one can expect the transitions $n\rightarrow n\pm2$
whenever we have a coupling of the gravitational field to a harmonic
oscillator \footnote{We will bring an intuitive understanding on the origin of the transition
$n\rightarrow n+2$. We can decompose the gravitational field into
the plane waves $\sim e^{-i(\omega_{k}t-kx)}$, and Taylor expand
in small displacements up to order $\mathcal{O}(x^{2})$: 
\begin{alignat}{1}
\hat{h}_{11}(t,x)\sim\hat{h}_{11}(t,0)+\frac{\partial\hat{h}_{11}(t,x)}{\partial x}\bigg\vert_{x=0}ikx\nonumber \\
-\frac{1}{2}\frac{\partial^{2}\hat{h}_{11}(t,x)}{\partial x^{2}}\bigg\vert_{x=0}k^{2}x^{2}+\cdots\label{eq:field}
\end{alignat}
where $k=\omega_{k}/c$ and $\omega_{k}$ is the angular frequency
of the gravitational field mode. The first term on the right-hand
side of Eq.~\eqref{eq:field} is a constant and can be omitted, while
the second linear term $\sim kx$ can be shown to vanish by considering
the Fermi Normal coordinates~\cite{Rakhmanov}, as a consequence
of the equivalence principle. From the remaining last term, we thus
find: 
\begin{equation}
\hat{h}_{11}(t,x)\sim-\frac{1}{2c^{2}}\frac{\partial^{2}\hat{h}_{11}(t,x)}{\partial t^{2}}\bigg\vert_{x=0}x^{2},\label{eq:field2}
\end{equation}
where we have used $\omega_{k}=kc$. From the gravitational coupling
to the matter component, $\hat{T}_{11}(x)\propto\delta(\hat{x}-x)$,
where $\hat{x}$ is the position of the harmonic oscillator, we find
the required quadratic coupling, i.e. $\hat{h}_{11}\hat{T}_{11}\propto\hat{x}^{2}$.
It is this matter-gravity coupling $\propto\hat{x}^{2}$
that leads to the transition $n\rightarrow n+2$. Since $\hat{x}\propto(\hat{a}+\hat{a}^{\dagger})$
and $\hat{x}\propto(\hat{b}+\hat{b}^{\dagger})$, we find the terms
$(\hat{a}^{\dagger})^{2}$and $(\hat{b}^{\dagger})^{2}$,
respectively. Combining these two terms we then get precisely the
term $(\hat{a}^{\dagger}\hat{b}^{\dagger})^{2}$ that we found using
the perturbation theory (see derivation below Eq.~\eqref{eq:Hab-2}).}. For example, it occurs also in the case of absorption/emission
of GWs of a specific polarisation ``+''~\citep{Toros}.

We now compute the reduced density matrix by tracing away system B
(we recall that our notation is $\vert n\rangle\vert N\rangle=\vert n\rangle_{A}\vert N\rangle_{B}$).
The concurrence in Eq.~(\ref{eq:concurrence2}) reduces to 
\begin{equation}
C\equiv\sqrt{2(1-\frac{1+(\mathfrak{g}/(2\omega_{\text{m}}))^{4}}{1+(\mathfrak{g}/(2\omega_{\text{m}}))^{2}})}\approx\sqrt{2}\frac{\mathfrak{g}}{\omega_{\text{m}}},\label{eq:concurrenceNewton-1-1-1}
\end{equation}
which is valid when the parameter ${\mathfrak{g}}/{\omega_{\text{m}}}\ll1$.
After inserting the coupling from Eq.~(\ref{eq:couplingGW-2}), we
find the concurrence to be: 
\begin{equation}
C=\frac{9\sqrt{2}G\hbar\omega_{\text{m}}}{16c^{4}d}.
\end{equation}
Note that the the degree of entanglement grows linearly with frequency
of the harmonic oscillators, does not depend on the mass, and scales
inversely with the distance between the two oscillators. The concurrence
in the case of the exchange of a graviton in the static limit dominates
over the non-static case, provided 
\begin{equation}
\omega_{\text{m}}^{2}<\frac{mc^{4}}{\hbar\omega_{\text{m}}d^{2}}.\label{threshold}
\end{equation}
In the original QGEM proposal~\cite{Bose:2017nin}, the proposed
inter separation distance between the two quantum superposition of
particles with mass $m\sim10^{-14}$kg is kept roughly at $d\sim100\times10^{-6}$m
in order to avoid Casimir induced entanglement~\cite{Bose:2017nin}.
If we wish to witness the entanglement in the non-static case, we
would require extremely high frequency oscillators (i.e., from Eqs.~\eqref{threshold}
we find $\omega_{{\rm m}}\gtrsim10^{21}$Hz), beyond the reach of
the current state of the art in a laboratory.

Let us highlight the link between LOCC/LOQC and the quantized graviton.
If the graviton were treated classically, then the final state of
the two harmonic oscillator states would have never evolved to an
entangled state like Eqs. (\ref{eq:finalMixed}) and (\ref{eq:finalGW})
-- in this case this amounts to $\hat{\gamma}_{11}$ and $\hat{\gamma}$
components. Indeed, a classical field is unable to give the operator-valued
shift of the vacuum energy in Eq.~\eqref{eq:dHGW} which led to the
quantum coupling in Eq.~\eqref{eq:Hab2} (i.e., a cross-product of
matter operators).


\section{Discussion\label{sec:Discussion}}

In this paper we have considered a specific example to reinforce the
importance of the quantum gravitational interaction in the QGEM protocol.
The crucial observation here is that the quantum nature of the gravitational
interaction yields operator valued shift in the gravitational Hamiltonians,
$\Delta\hat{H}_{g}$, see Eq.(\ref{eq:Newton},\ref{eq:dH5}). Classical
gravity will only yield a real-valued shift in $\Delta H_{g}$.

In particular, we considered the two quantum harmonic oscillators
separated by a distance $d$ interacting via the exchange of a graviton
comprising of the spin-2 and spin-0 components. We have shown that
the quantum nature of the graviton (for both spin-2 and spin-0, $\hat{h}_{00}\equiv\hat{\gamma}_{00}+(1/2)\hat{\gamma}$)
is essential to create an entangled state with the ground and excited
states of the harmonic oscillators forming the Schmidt basis.

Similar physics arises in the non-static case as well. The quantum
nature of the graviton (i.e., $\hat{h}_{01}\equiv\hat{\gamma}_{01}$
component) will generate a two-mode squeezed state of the two harmonic
oscillators, Eq.~\eqref{eq:finalMixed}. On the other hand, $\hat{h}_{11}\equiv\hat{\gamma}_{11}-(1/2)\hat{\gamma}$
component is crucial to entangle with the ground and the second excited
states of the harmonic oscillators. It is also interesting to note
that these latter states, Eq.~\eqref{eq:finalGW}, have never been
presented previously, to our knowledge, in any context in the vast
literature on entangled harmonic oscillators (for example in the quantum
optics or in the allied literature). They are particular to the nature
of spin-2 graviton.

We have obtained all the results relying only on the elementary perturbation
theory; the wave function was evaluated up to the first order, and
the correction to the graviton vacuum was computed up to the second
order (to obtain the non-vanishing contribution to the vacuum energy).
Both the wave function calculations and the correction to the energy
of the vacuum suggest that the \emph{quantum interaction between the
graviton and the matter} is crucial to obtain entanglement, reinforcing
that the LOCC can not yield or lead to the increment in the entanglement~\footnote{If one limits the discussion to the non-relativistic models of gravity
and simply postulates the interaction term as $\sim1/\vert\hat{\bm{r}}_{a}-\hat{\bm{r}}_{b}\vert$
one cannot say much about the underlying dynamical degrees of freedom
of the gravitation field. Here, we have shown that in the perturbative
canonical quantum theory of gravity we can account for the dynamical
degrees of freedom. These are crucial to obtain the correct shift in the operator
valued gravitational energy which give rise to the quantum matter-matter
interaction. Other theories beyond GR would require a similar analysis
of the dynamical degrees of freedom of the gravitational field.}.

We computed the entanglement concurrence and showed that the concurrence
is always positive for a quantum gravitational field (indicating entanglement),
but would remain zero for a classical gravitational field (no entanglement).
Moreover, the entanglement can be regarded as due to the operator
valued shifts of the vacuum energy.

So far we have kept our investigation limited to the local quantum
interaction between matter and the gravitational field -- our $\hat{H}_{\text{int}}$
was strictly local. It would be interesting to study what would happen
if the locality in the gravitational interaction is abandoned~\citep{Tomboulis,Biswas:2005qr,Biswas:2011ar,Modesto,Marshman:2019sne}.
Giving up local gravitational interaction will help us to further
investigate the entanglement in theories beyond GR, and in quantum
theories of gravity where non-local interactions enters in various
manifestations, see~\citep{Kiefer,deLacroix:2017lif,Loll:2019rdj,Sorkin:2007qi}.
We can also attempt to compute the entanglement by modifying the graviton
propagator in a non-perturbative formulations of quantum gravity~\cite{Rovelli:2005yj,Bianchi:2006uf}.
Similar computations to the entanglement can be computed within perturbative
quantum gravity but with higher post Newtonian Hamiltonians in $3+1$
dimensions, see~\cite{Bern:2019nnu,Bini:2020wpo}.

In summary, our results corroborate the importance of the QGEM experiment,
which relies on the fact that the two quantum superposed masses kept
at a distance can entangle via the quantum nature of the graviton.
This would be crucial in unveiling the quantum properties of the spin-2
graviton which is hitherto a hypothetical particle responsible for
the fluctuations of the space-time in the context of a perturbative
quantum gravity.

\section*{Acknowledgments}

MT and SB would like to acknowledge EPSRC grant No.EP/N031105/1, SB
the EPSRC grant EP/S000267/1, and MT funding by the Leverhulme Trust
(RPG-2020-197). MS is supported by the Fundamentals of the Universe
research program within the University of Groningen. AM's research
is funded by the Netherlands Organisation for Science and Research
(NWO) grant number 680-91-119.


\end{document}